\documentclass[preprint,aps,amsmath,nofootinbib,tightenlines,floatfix]{revtex4}
\usepackage{graphicx, epsfig}
\usepackage{multirow}

\begin{document}
\title{$h\rightarrow \mu^+ \mu^-$ via $t\bar{t}h$ Production at the LHC}
\author{Shufang Su\footnote{shufang@physics.arizona.edu} \,and\, Brooks Thomas\footnote{brooks@physics.arizona.edu}}
\affiliation{Department of Physics, University of Arizona, Tucson, AZ  85721 USA}

\begin{abstract}
In this work, we examine the process at the LHC in which a Higgs boson is produced in 
association with a  $t\bar{t}$ pair and subsequently decays to a 
pair of muons.    We show that the statistical significance for the discovery of a 
light, Standard-Model Higgs boson with a mass around 120~GeV in this 
channel is comparable to those for other processes (gluon fusion, weak-boson fusion) in
which the Higgs decays to a muon pair. 
Combining all three of these channels,  we show that 
evidence for a Higgs boson with a mass in the range 
$115\mathrm{~GeV}< m_h< 130\mathrm{~GeV}$ could be obtained at the $3\sigma$ significance 
level with an integrated luminosity of ${300\ {\rm fb^{-1}}}$.  
We also calculate the enhancement factor to the cross-section that would be needed 
to discover a non-standard Higgs boson in this channel.
\end{abstract}

\maketitle

\newcommand{\newc}{\newcommand}
\newc{\gsim}{\lower.7ex\hbox{$\;\stackrel{\textstyle>}{\sim}\;$}}
\newc{\lsim}{\lower.7ex\hbox{$\;\stackrel{\textstyle<}{\sim}\;$}}

\def\beq{\begin{equation}}
\def\eeq{\end{equation}}
\def\beqn{\begin{eqnarray}}
\def\eeqn{\end{eqnarray}}
\def\calM{{\cal M}}
\def\calV{{\cal V}}
\def\calF{{\cal F}}
\def\half{{\textstyle{1\over 2}}}
\def\quarter{{\textstyle{1\over 4}}}
\def\ie{{\it i.e.}\/}
\def\eg{{\it e.g.}\/}
\def\etc{{\it etc}.\/}


\def\inbar{\,\vrule height1.5ex width.4pt depth0pt}
\def\IR{\relax{\rm I\kern-.18em R}}
 \font\cmss=cmss10 \font\cmsss=cmss10 at 7pt
\def\IQ{\relax{\rm I\kern-.18em Q}}
\def\IZ{\relax\ifmmode\mathchoice
 {\hbox{\cmss Z\kern-.4em Z}}{\hbox{\cmss Z\kern-.4em Z}}
 {\lower.9pt\hbox{\cmsss Z\kern-.4em Z}}
 {\lower1.2pt\hbox{\cmsss Z\kern-.4em Z}}\else{\cmss Z\kern-.4em Z}\fi}




\section{Introduction}

The CERN Large Hadron Collider (LHC) --- a $pp$
collider with center-of-mass energy $\sqrt{s}=14$~TeV --- will begin taking
data in the very near future.  One of the primary missions of the 
LHC is to investigate the physics behind electroweak symmetry-breaking.
Since many of the most attractive candidate theories involve a Higgs sector
containing one or more light, scalar degrees of freedom, a great deal of 
effort has been invested into determining how best to
search for these scalars (and in particular the lightest CP-even scalar) at the LHC 
(see~\cite{ATLASTDR,CMSTDR,AsaiVector,CMSHiggs}).    
For a relatively light Higgs boson in the Standard Model (SM), with $114\mbox{~GeV}\lesssim m_h\lesssim 125\mbox{~GeV}$, 
the most favored channels are $gg\rightarrow h\rightarrow\gamma\gamma$, $t\bar{t}h(h\rightarrow b\bar{b})$.  
For a Higgs with intermediate mass $125\mbox{~GeV}\lesssim m_h\lesssim 140\mbox{~GeV}$, the most promising channel is the weak-boson fusion (WBF)~\cite{wbf} process $qq'\rightarrow qq'h(h\rightarrow\tau\tau)$~\cite{RainwaterHtautau}.  
For a heavier Higgs, with  
$ m_h\gtrsim 140\mbox{~GeV}$, the channels of interest are $h\rightarrow WW^{\ast}$ 
and $h\rightarrow ZZ^{\ast}$, with the Higgs produced through either gluon fusion or 
WBF~\cite{AsaiVector,CMSHiggs}.

Clearly, additional channels which might provide evidence for a light, CP-even Higgs scalar in 
these mass ranges can also play a crucial role in Higgs phenomenology.
This is especially true if the Higgs sector realized in nature turns out to be more 
complicated than the single $SU(2)$ doublet of the SM.  In such cases, 
the most promising discovery channels 
could turn out to be quite different from those pertinent to a SM Higgs 
boson~\cite{UnusualHiggs}.  For this reason, it has become increasingly clear that 
channels for which the significance of discovery for a Standard Model Higgs is low may be of crucial importance for detecting whatever variety of Higgs boson is
actually out there --- provided of course that nature employs a scalar Higgs sector to
break the electroweak symmetry.   
Furthermore, if a light, CP-even Higgs boson is discovered early at the LHC, within the 
first $10-30~\mathrm{fb}^{-1}$ of data, processes which can put the most
stringent bounds on parameters such as its mass, its couplings to the Standard 
Model fermions, etc.\ will become increasingly important for precision analyses
of the Higgs sector that could point the way toward new physics.

One Higgs-decay channel that is of particular interest is $h\rightarrow \mu^+\mu^-$.  The primary advantage of this channel is that a signal involving a pair of high-$p_{T}$ muons will be easy to identify at the LHC.  Indeed, the muon-identification efficiency at the LHC detectors is more that 90\%\cite{ATLASTDR, CMSTDR}.  Once a Higgs boson is discovered in this channel, its mass could be readily reconstructed with high precision.  Additionally, such a channel could be used in determining the muon Yukawa coupling.  This is of particular interest because most of the existing literature on the measurement of Higgs-boson Yukawa couplings focuses on the third generation Yukawa couplings, $y_b$, $y_\tau$ and $y_t$.  Consequently, $h\rightarrow\mu^+\mu^-$ processes could be important for determining whether or not the effective Higgs Yukawa couplings are indeed generation-universal.

The difficulty with $h\rightarrow \mu^+\mu^-$ processes is the small Higgs branching 
fraction into muon pairs, both in the Standard Model --- in which it is of $\mathcal{O}(10^{-4})$ --- and in most simple extensions of the Higgs sector.    
Indeed, under the assumption of universal Yukawa couplings among the lepton generations, the small size of the muon mass $m_{\mu}$ compared to $m_\tau$ results in 
${\rm BR}(h\rightarrow \mu\mu)$ being roughly two orders of magnitude smaller than 
${\rm BR}(h\rightarrow \tau\tau)$.  Consequently, while a great deal of attention has been
focussed on processes in which the Higgs boson decays to a tau  pair 
(for example, weak-boson fusion with $h\rightarrow \tau^+\tau^-$ is 
now regarded as one of the promising 
discovery channel for the SM Higgs in the intermediate mass 
region~\cite{wbf, ATLASTDR, CMSTDR}), $h\rightarrow\mu^+\mu^-$ processes have not been 
extensively considered.  Nevertheless, some parton-level studies have been carried out:
investigations of $gg\rightarrow h\rightarrow \mu\mu$ and the weak-boson fusion process 
$qq'\rightarrow qq'h(h\rightarrow \mu\mu)$ at the LHC were performed in Ref.~\cite{HanMcElrath} and Ref.~\cite{PlehnRainwater}, respectively.  It was found that by combining the 
results from both of these channels, an observation of $h\rightarrow\mu^+\mu^-$ at the 
$3\sigma$ level could be obtained with $300\ {\rm fb}^{-1}$ of integrated luminosity for low Higgs masses. 
Higgs decays to muon pairs in the Minimal Supersymmetric Standard Model have also been 
studied (see, for example, Ref.~\cite{BargerKao}).

In this letter, we examine the associated Higgs production process $t\bar{t}h$ 
with $h\rightarrow\mu^+\mu^-$ at the LHC.  While the rate for 
this process is far smaller than that for the      
$t\bar{t}h(h\rightarrow\tau\tau)$ channel discussed in~\cite{BelyaevReina}, a muonic final
state has a number of advantages.  Perhaps the most significant of these is that the final-state muon pair affords an exceptionally good Higgs-mass resolution --- of similar order (around 1\%) to that afforded by $gg\rightarrow h\rightarrow \gamma\gamma$~\cite{MassRes2Phot,CMSTDR}.
In addition, the absence of any additional sources of missing energy for semileptonic top
decays (beyond the single neutrino from $t\rightarrow b\ell\nu$) reduces the uncertainty in
top-quark reconstruction.  We show that, for a SM Higgs boson with a mass around 
120~GeV, the statistical 
significance of discovery in the $t\bar{t}h(h\rightarrow\mu\mu)$ channel is on the same order
as that obtained from the gluon fusion and WBF channels.  We also investigate the prospects for detecting a SM
Higgs boson via its decays to muon pairs using the combined gluon fusion, WBF, and $t\bar{t}h$ channels, and discuss the potential implications 
for Higgs discovery using the $t\bar{t}h(h\rightarrow \mu\mu)$ channel in
beyond-the-Standard-Model scenarios.
    

\section{Signal and Backgrounds}
 
For a SM Higgs sector, the $t\bar{t}h(h\rightarrow\mu\mu)$ channel will be
useful primarily in the case where the Higgs boson is light: 
$114\mbox{~GeV}\lesssim m_h\lesssim 140\mbox{~GeV}$.  (The production cross-sections drop quickly for heavier Higgs masses.)  Here, we present results 
for the specific choices $m_h=115$~GeV, 120~GeV, 130~GeV, and 140~GeV. 
The leading-order (LO) $t\bar{t}h$ production
cross-section and Higgs branching fraction to muons 
for each of these choices are listed in Table~\ref{tab:SigBR}.
The former were calculated using the MadGraph package~\cite{MadGraph}, with 
factorization and renormalization scales both set to $(2m_t+m_h)/2$;
the latter were calculated using HDECAY~\cite{HDECAY}.
It is apparent from Table~\ref{tab:SigBR} that the leading-order (LO) $t\bar{t}h$ 
production cross-section for a Higgs boson in this mass range at the LHC ranges 
from around 350 to 600~fb, while the Higgs branching fraction to muons
is of ${\mathcal{O}}(10^{-4})$, as discussed above.  

Since this implies a rather small 
overall rate for $t\bar{t}h(h\rightarrow\mu\mu)$ events, we would like to include in 
our analysis all final states which permit an unambiguous reconstruction of both $t$ 
and $\bar{t}$ from their decay products.  These include processes in which both $W$ 
bosons  decay hadronically (45.7\% of the time) and semileptonic decays in which 
the charged lepton is either an electron or a muon (28.8\% of the time).  
Therefore, we study two types of signals:
\renewcommand{\theenumi}{\alpha{enumiii}}
\begin{itemize}
\item[(I)] hadronic signatures: 6 jets  
$+$ $\mu^+\mu^-$,
\item[(II)] semileptonic signatures: 4 jets 
$+$ $\mu^+\mu^- \ell^{\pm}$ $+$ missing $E_T$, $\ell=e,\mu$.
\end{itemize}
For fully-leptonic $t\bar{t}$ decay, the masses of the
top quarks cannot be completely reconstructed, due to the presence of multiple 
neutrinos in the final state.  We therefore do not consider this final state in 
our results, but note that a more detailed analysis based on kinematical distributions 
as in the top-mass reconstruction methods of~\cite{ATLASTDR,TopRecon} could perhaps 
render this channel useful.

It may be noted that no distinction between $b$-jets and other jets was made in the
definition of signals (I) and (II). 
Indeed, $b$-tagging is not used in this analysis and that the cuts outlined below are 
applied to all jets, irrespective of their $b$-character.  The reason for this is that 
the $b$-tagging efficiency at ATLAS is
around 50\%~\cite{ATLASTDR} (the efficiency at CMS is comparable), which would lead 
to a substantial loss in
signal events.  However, we shall justify this procedure below and show that the 
relevant reducible backgrounds 
can be effectively eliminated merely by demanding that $t$, $\bar{t}$, $h$ and 
both $W$-bosons, can all be 
appropriately reconstructed from the momenta of the final-state particles.          
      
 \begin{table}[thb!]
\begin{center}
  \begin{tabular}{|ccc|}\hline
   $m_h$ (GeV) & $\sigma_{t\bar{t}h}$ (fb) & {BR}$(h\rightarrow\mu\mu) $ \\  \hline
    115 & 606.1 & $2.57\times10^{-4}$ \\
    120 & 538.1 & $2.40\times10^{-4}$ \\
    130 & 430.8 & $1.89\times10^{-4}$ \\  
    140 & 347.5 & $1.24\times10^{-4}$ \\ \hline   
   \end{tabular}
\end{center}
 \caption{Leading-order (LO) SM $t\bar{t}h$ production cross sections at the LHC and Higgs branching ratios to muons for
several different values of the Higgs mass $m_h$.}
\label{tab:SigBR}
 \end{table}

The effect of next-to-leading order (NLO) QCD corrections is incorporated in our analysis 
in the standard manner, via the introduction of a $K$-factor for each relevant signal
or background process.  NLO corrections to the $t\bar{t}h$ production cross-section
were calculated in Ref.~\cite{NLOforttH} for $m_h=120$ GeV.  For this choice of Higgs mass,
the $K$-factor was found to be about 1.2 at the scale $\mu=(2m_t+m_h)/2$, 
and was found not to vary
substantially with $m_h$ over the mass range considered here.  
We therefore take $K_S$, the $K$-factor for the signal process, to be
equal to 1.2 for all $115\mbox{~GeV}\leq m_h \leq 140\mbox{~GeV}$. 

The primary irreducible Standard Model background for all final states resulting 
from $t\bar{t}h(h\rightarrow\mu\mu)$ comes from similar processes in which the Higgs boson 
is replaced by an off-shell photon $\gamma^{\ast}$ or a $Z$ boson.  The background was also calculated
using MadGraph~\cite{MadGraph}, with both the factorization and renormalization scales set 
to $(2m_t+m_Z)/2$.  We found the tree-level cross section for $t\bar{t}Z/\gamma^*(Z/\gamma^{\ast}\rightarrow \mu\mu)$ to be 32.27~fb, where 
we require $p_T>10$~GeV for  
final-state muons to keep the results well-behaved when soft photons are taken 
into account. 
This result is also modified at next-to-leading order by a $K$-factor, which was
calculated in Ref.~\cite{NLOforttZ} and found to be $K_{BG}=1.35$ 
at the scale $\mu=(2m_t+m_Z)/2$.
Reducible backgrounds such as $bbWWZ$, 
also exist, but they can be eliminated by a sensible choice
of cuts.  In particular, mandating the successful reconstruction of the top quarks in
both the hadronic channel and the semileptonic channel
will render such backgrounds negligible.  Therefore, in our analysis below, we only consider the irreducible background coming from $t\bar{t}Z/\gamma^{\ast}$.

We impose two sets of cuts 
on the signal and background data for each type of signature 
(hadronic or semileptonic) under consideration.  
The first of these sets (which we will refer to as the Level~I cuts) is applied
universally to all processes, and is designed to reproduce a realistic detector acceptance:
\begin{itemize}
  \item $p_T^{\ell}>20$~GeV  for all leptons, $|\eta_{\mu}|<2.4$, $|\eta_{e}|<2.5$,
  \item $p_T^{j}>15$~GeV, $|\eta_{j}|<3.0$ for all jets (including $b$ jets),
  \item $\Delta R_{\ell j}>0.4$, $\Delta R_{jj}>0.5$, $\Delta R_{\ell\mu}>0.5$.
\end{itemize} 
Here, $\Delta R_{ab} =\sqrt{(\Delta\phi_{ab})^2+(\Delta\eta_{ab})^2}$ denotes the ``lego-plot'' separation distance
between final-state particles $a$ and $b$, and $p_T^a$ denotes the transverse momentum of $a$. 

Reconstruction requirements underlie the second set of cuts, the specifics of which depend on the particular final states under consideration.  
Since the invariant mass resolution for the muon 
pair will be quite good at both ATLAS~\cite{ATLASTDR} and CMS~\cite{CMSTDR}, we retain only events in which the invariant 
mass of some pair of muons lies within the range $|M_{\mu\mu}-m_h|<2.5$~GeV for the particular value of $m_h$ under 
consideration.  In addition to this requirement, for the fully hadronic final state (I), which involves 
6~jets~$+$~$\mu^+\mu^-$, we demand that some combination of jet momenta exists for which the invariant masses $M_{j_aj_bj_c}^2=(p_{j_a}+p_{j_c}+p_{j_c})^2$ of two different sets of jets $(j_a,j_b,j_c)$ both reconstruct a top quark.  Furthermore, we require that within each such set, one combination of jet momenta $M_{j_aj_b}^2=(p_{j_a}+p_{j_b})^2$ reconstructs a $W$ boson.  
Thus the full roster of Level~II cuts in the fully hadronic channel comprises:
\begin{itemize}
  \item 6 jets and 2 muons passing the Level~I cuts, 
  \item two groups of three jets each for which $|M_{\mathit{j_aj_bj_c}}-m_t|<50$~GeV,
  \item one jet pair within each such group for which $|M_{\mathit{j_aj_b}}-M_W|<40$~GeV,
  \item $|M_{\mu\mu}-m_h|<2.5$~GeV.
\end{itemize} 

For the semileptonic signature (II), which involves 
4~jets~$+$~$\mu^+\mu^- \ell^{\pm}$~$+$~missing $E_T$, 
the situation is complicated both by
the presence of missing transverse momentum from the neutrino and by combinatorial issues that arise when $\ell^{\pm}$ is a muon.  
To address the former, we follow the standard procedure~\cite{ATLASTDR}, which is to assume that a single neutrino produced by the 
leptonically-decaying top is responsible for the entirety of the missing momentum in the transverse plane and that
$M_W^2=(p_{\nu}+p_{\ell})^2$.  Under this assumption, one can solve for the longitudinal component $p_{z}^{\nu}$ of the neutrino 
momentum up to a sign ambiguity.  
Of the two resulting solutions for $p_z^\nu$, we select the one with the larger absolute value 
and use it to reconstruct a mass for the leptonically-decaying top.  
For final states involving at least one electron, 
we then require that there exist one jet and one electron which, in this manner, reconstruct $m_t$ to within 50~GeV.  
For final states that contain at least three muons,
we accept any event for which there exists 
some combination of muons which
reconstructs both $m_h$ and $m_t$ successfully according to the method outlined above, 
provided that the muons used to reconstruct the Higgs mass have opposite sign.  However since combinatorial issues of this sort can have a pronounced effect when the primary background is associated with the tail of a kinematical distribution, far away from the $Z$-pole, we apply a more stringent constraint in this case and require that $m_t$ be reconstructed within 10~GeV.     
To reconstruct the remaining top quark, we use the same method employed in the fully-hadronic case: we require that there exist one combination of three other jets whose invariant mass reconstructs $m_t$ within 50~GeV, two of which reconstruct $M_W$ within 40~GeV.    

To recapitulate, then, the Level~II cuts
for the semileptonic case are:          
\begin{itemize}
  \item 4 jets and 3 charged leptons (at least two of which are muons) passing the Level~I cuts, 
  \item a groups of three jets each for which $|M_{\mathit{j_aj_bj_c}}-m_t|<50$~GeV,
  \item one jet pair within this group for which $|M_{\mathit{j_aj_b}}-M_W|<40$~GeV,
  \item 2 opposite-sign muons for which $|M_{\mu\mu}-m_h|<2.5$~GeV,
  \item one additional electron and one additional jet for which $|M_{\mathit{je\nu}}-m_t|<50$~GeV, or one additional muon and one additional jet for which $|M_{\mathit{j\mu\nu}}-m_t|<10$~GeV,
\end{itemize}
where the neutrino four-momentum is reconstructed in the manner discussed above.  
Note that the process in which the charged lepton from the leptonically-decaying top 
is a muon and the process in which it is an electron represent two distinct channels with different backgrounds, detection efficiencies, etc.
These results pertaining to these channels will thus be reported separately. 

\section{Results}
 
The results of our analysis of $t\bar{t}h(h\rightarrow\mu\mu)$ in the hadronic and semileptonic channels 
are displayed in Tables~\ref{tab:had} and~\ref{tab:semi}, respectively.  The respective
efficiencies $\epsilon_S$ and $\epsilon_B$ of the cuts discussed above in reducing the number 
of signal and background events in each case are shown, along with cross-sections for 
the corresponding processes which include the effect of these efficiencies.  Note that 
here we have defined $\epsilon_S$ and $\epsilon_B$ for each channel to include the branching
fraction of $t\bar{t}$ to the corresponding final state.  In addition, 
signal-to-background ratios $S/B$ and statistical significances ($S/\sqrt{S+B}$) are 
also displayed.  Signal and background 
events for each process were generated using the MadEvent package~\cite{MadGraph} and processed using~PYTHIA~\cite{PYTHIA}.  The generated events were then passed through PGS4~\cite{pgs} for detector simulation.
A total of $40,000$ events were simulated for the signal process; for the 
background process, on which the effect
of the cuts is more severe, ten times that number were simulated in order to 
limit the effect of 
statistical fluctuations.   

\begin{table}
\begin{center}
\begin{tabular}{|c|cc|cc|c|c|} \hline
\multirow{2}{*}{~$m_h$ (GeV)~} &\multirow{2}{*}{$\epsilon_S$} & 
\multirow{2}{*}{~$\sigma_S$ (ab)~} &
\multirow{2}{*}{$\epsilon_{B}$} &\multirow{2}{*}{~$\sigma_B$ (ab)~}& \multirow{2}{*}{~$S/B$~}&$~S/\sqrt{S+B}~$ \\
&&&&&&${\mathcal{L}}=300\ {\rm fb}^{-1}$\\ \hline
    115 & ~8.1\%~ &~15.1~& ~0.028\%~ & ~12.3~ & 1.22 & 2.23 \\  
    120 & ~7.9\%~ &~12.2~& ~0.023\%~ & ~10.0~ & 1.21 & 2.00 \\ 
    130 & ~8.2\%~  &~8.0~& ~0.017\%~ & ~7.3~ & 1.10 & 1.59 \\ 
    140 & ~8.2\%~  &~4.2~& ~0.015\%~ & ~6.6~ & 0.64 & 0.99 \\ \hline
\end{tabular}
\end{center}
\caption{Signal and background results for $t\bar{t}h(h\rightarrow\mu\mu)$ in the fully hadronic channel for several different values of $m_h$.   
The signal and background efficiencies $\epsilon_S$ and $\epsilon_B$ (including the top quark decay branching ratios) associated with the combined Level~I and Level~II cuts described in the text are displayed here, along with the signal and background cross-sections (in ab) after all cuts have been applied.  The cross sections quoted here include the relevant NLO $K$-factors for both signal and background processes.  The statistical significance $S/\sqrt{S+B}$ is also given for $300\ {\rm fb}^{-1}$ of integrated luminosity at both ATLAS and CMS, along with the result for $S/B$.}
\label{tab:had}
 \end{table}

The hadronic channel gives the best statistical significance for $t\bar{t}h(h\rightarrow \mu\mu)$.  We find that evidence for a light ($m_h=115$~GeV) SM Higgs boson may be obtained at the $2.23\sigma$ significance level with $300\ {\rm fb}^{-1}$ of integrated luminosity at both ATLAS and CMS.  The statistical significance of this channel decreases with increasing Higgs mass, primarily due to the decrease both in the $t\bar{t}h$ production cross-section and in the branching ratio of Higgs decay to muons.  This notwithstanding, for all $115{\mathrm{~GeV}}<m_h< 140$ GeV, the signal cross-section for this process is large enough that a reasonable number of signal events can be expected at such luminosities. 

\begin{table}
\begin{center}
\begin{tabular}{|c|ccccc|ccccc|c|} \hline
\multirow{2}{*}{\parbox{1.15cm}{$m_h$ \\(GeV)}}& \multicolumn{5}{c|}{$W\rightarrow e\nu_e$} & 
\multicolumn{5}{c|}{$W\rightarrow \mu\nu_\mu$} & 
\multirow{2}{*}{\parbox{2.5cm}{$S/\sqrt{S+B}$\\ (${\mathcal{L}}=300\ {\rm fb}^{-1}$)}} \\
\cline{2-11} 
& {$\epsilon_S$} & {$\sigma_S {\rm (ab)}$} &
 {$\epsilon_{B}$} & {$\sigma_B {\rm (ab)}$} & $~S/B~$ &
 {$\epsilon_S$} & $\sigma_S {\rm (ab)}$ &
 {$\epsilon_{B}$} & {$\sigma_B {\rm (ab)}$} &
 $~S/B~$& \\ \hline
    115 & 1.6\% & 3.0 & 0.0045\% & 1.9 & 1.55 & 0.74\% & 1.4 & 0.011\% & 5.0 & 
         0.28 & 1.13  \\  
    120 & 1.8\% & 2.7 & 0.0025\% & 1.1 & 2.51 & 0.70\% & 1.1 & 0.013\% & 5.4 & 
         0.20 & 1.13  \\ 
    130 & 1.7\% & 1.7 & 0.0030\% & 1.3 & 1.26 & 0.76\% & 0.7 & 0.010\% & 4.4 & 
         0.17 & 0.79  \\ 
    140 & 1.7\% & 0.9 & 0.0013\% & 0.5 & 1.63 & 0.74\% & 0.4 & 0.008\% & 3.4 & 
         0.11 & 0.59  \\ \hline
\end{tabular}
\end{center}
 \caption{Signal and background results for $t\bar{t}h(h\rightarrow\mu\mu)$, with 
$t\bar{t}\rightarrow 2$ jets+$\ell^{\pm}+ $missing $E_T$, for both $\ell=e$ 
and $\ell=\mu$, are shown here for several different values of $m_h$, including all 
relevant $K$-factors, etc.  For further explanation of the notation, see the caption 
for Table~\ref{tab:had}.  The combined statistical significance $S/\sqrt{S+B}$ in 
the semileptonic channel from both contributions for $\mathcal{L}=300\ {\rm fb}^{-1}$ 
is also supplied.}
\label{tab:semi}
 \end{table}

By contrast, the semileptonic channels are slightly less promising, primarily because the efficiency factor $\epsilon_S$ is significantly smaller than in the fully-hadronic channel.  Furthermore, a much larger fraction of background events survive the cuts in the channel in which the charged lepton produced by the leptonically-decaying top is a muon than in the channel where it is an electron. 
Taking into account 
the contributions from both of these final states, we find that with 300~${\rm fb}^{-1}$ of
integrated luminosity at both ATLAS and CMS, semileptonic $t\bar{t}h$ events can provide evidence at the $1.13\sigma$ significance level for a light ($m_h=115$~GeV) Higgs boson.  Since the signal cross-sections are relatively small for both contributing channels, however, with $\sigma_S$ of $\mathcal{O}(ab)$ after all cuts have been applied, only a small number of surviving events will be produced at such luminosities --- at least for a standard model Higgs boson.   

Table~\ref{tab:combine} presented the $t\bar{t}h$ results combining all three channels.  We also list the results for $gg\rightarrow{h}\rightarrow \mu\mu$~\cite{HanMcElrath} and $qq'\rightarrow qq'h(h\rightarrow\mu\mu)$~\cite{PlehnRainwater} for comparison.  
For a SM Higgs mass around 120~GeV, the statistical significance in the
$t\bar{t}h(h\rightarrow \mu\mu)$ channel is similar to or even higher than
that in the other two channels.  With
$\mathcal{L}=300\ {\rm fb}^{-1}$, we observe that a statistical significance of 
$4.19\sigma$ may be obtained via the combined channels.  This represents a substantial
increase over the $3.37\sigma$ obtained from considering the gluon-fusion and WBF processes alone.  In Table~\ref{tab:luminosity}, we presented the integrated luminosity at the LHC that is needed for a $3\sigma$ and $5\sigma$ discovery with $t\bar{t}h(h\rightarrow \mu\mu)$ alone (combining both the hadronic channel and the semileptonic channels), and with all three production processes ($t\bar{t}h$, gluon fusion, WBF) included.  For a Higgs boson just slightly
heavier than the LEP bound of 114~GeV~\cite{LEPBound}, a $3\sigma$ discovery can be obtained with
around $150\ {\rm fb}^{-1}$ of integrated luminosity.  

\begin{table}
\begin{tabular}{|c|ccc|cc|}\hline
\multirow{2}{*}{~$m_h$ (GeV)~}& \multicolumn{3}{c|}{$S/\sqrt{S+B}$} & \multicolumn{2}{c|}{~$S/\sqrt{S+B}$~} \\
& $t\bar{t}h$ & ~~GF~~ & ~WBF~ &~(GF+WBF)~& ~Combined~\\
\hline
115 & 2.50 & 2.41 & 2.35 & 3.37 & 4.19 \\
120 & 2.30 & 2.51 & 2.37 & 3.45 & 4.15 \\
130 & 1.77 & 2.25 & 2.25 & 3.18 & 3.64 \\
140 & 1.16 & 1.61 & 1.58 & 2.26 & 2.54 \\ \hline
\end{tabular}
\caption{Combined statistical significance for $t\bar{t}h(h\rightarrow\mu\mu)$ in the Standard Model, displayed alongside those for gluon fusion~\cite{HanMcElrath} and weak-boson fusion~\cite{PlehnRainwater} processes in which the Higgs boson decays to $\mu^+\mu^-$.  The significance values quoted here correspond to $\mathcal{L}=300\ {\rm fb}^{-1}$ for both ATLAS and CMS.}
\label{tab:combine}
\end{table}

\begin{table}
\begin{tabular}{|c|cc|cc|}\hline
\multirow{2}{*}{~$m_h$(GeV)~} &
\multicolumn{2}{c|}{$t\bar{t}h$} &
\multicolumn{2}{c|}{$t\bar{t}h+$GF$+$WBF} \\ &
~${\mathcal{L}}^{3\sigma}$ $({\rm fb}^{-1})$~ &
~${\mathcal{L}}^{5\sigma}$ $({\rm fb}^{-1})$~ & 
~${\mathcal{L}}^{3\sigma}$ $({\rm fb}^{-1})$~ &
~${\mathcal{L}}^{5\sigma}$ $({\rm fb}^{-1})$~ \\ \hline
115&  432 & 1200 & 154 &  427 \\
120&  511 & 1419 & 157 &  436 \\
130&  859 & 2386 & 204 &  566 \\
140& 2014 & 5594 & 420 & 1166 \\ \hline
\end{tabular}
\caption{The integrated luminosity needed to claim a $3\sigma$ or $5\sigma$ discovery of the Standard Model Higgs boson at the LHC in the $t\bar{t}h(h\rightarrow\mu\mu)$ channel for a
several different choices of $m_h$.  The luminosity needed to claim a $3\sigma$ or $5\sigma$ discovery using the combined gluon fusion, weak-boson fusion, and $t\bar{t}h$ Higgs production channels, with $h\rightarrow\mu^+\mu^-$, is also shown.}
\label{tab:luminosity}
\end{table}

\begin{figure}[thb!]
\centerline{
   \epsfxsize 3.0 truein \epsfbox {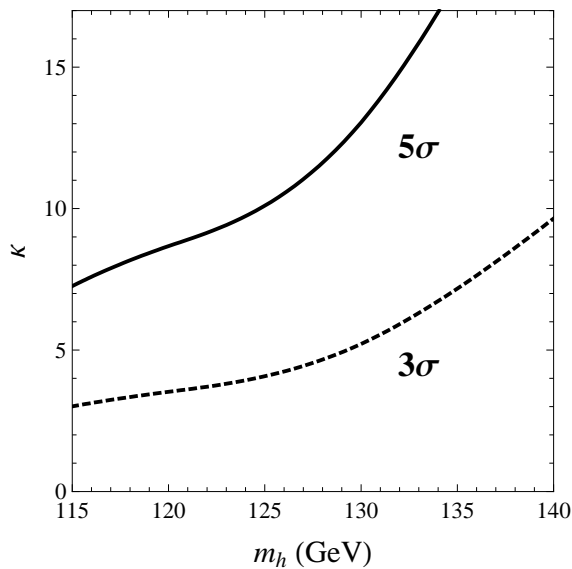} }
\caption{Plot of the enhancement factor $\kappa$ --- see Eq.~(\ref{eq:kappadef}) --- 
necessary to obtain evidence for a light, non-standard CP-even Higgs boson at the $3\sigma$ or $5\sigma$ level at LHC with $\mathcal{L}=100\ {\rm fb}^{-1}$, given as a function of the Higgs boson mass $m_h$.  The dotted line corresponds to a $3\sigma$ effect, whereas the solid line corresponds to a $5\sigma$ effect.
\label{fig:Kappas}}
\end{figure}  

We have shown that the $t\bar{t}h(h\rightarrow\mu\mu)$ channel can provide $2.5\sigma$ evidence for a light, SM Higgs boson around 115~GeV.  In non-Standard Model Higgs scenarios, however, the branching ratio for $h\rightarrow\mu^+\mu^-$ and/or $t\bar{t}h$ production cross section might potentially be enhanced, rendering this channel even more significant for Higgs-boson detection.  Depending on the enhancement factor, a $5\sigma$ discovery might even possible --- and at a reasonably low integrated luminosity.  In order to quantify these effects, we define an overall enhancement factor $\kappa$, which represents the net effect of all modifications to ${\rm BR}(h\rightarrow\mu\mu)$ and the  $t\bar{t}h$ production cross-section in a given extension of the SM Higgs sector:      
\begin{equation}
\kappa\equiv\frac{[\sigma(pp\rightarrow t\bar{t}h)\times {\rm BR}(h\rightarrow \mu\mu)]_{NP}}
{[\sigma(pp\rightarrow t\bar{t}h)\times {\rm BR}(h\rightarrow \mu\mu)]_{\rm SM}}.
\label{eq:kappadef}
\end{equation}

Figure.~\ref{fig:Kappas} shows the enhancement factor $\kappa$ as a function of Higgs mass for a $3\sigma$ or $5\sigma$ discovery at the LHC with $100\ {\rm fb}^{-1}$ of integrated luminosity.  For values of $m_h$ in the range $115 \leq m_h \lesssim 130$ GeV, an enhancement factor of
$\kappa\gtrsim 3-5$ (depending on $m_h$) is sufficient to result in a $3\sigma$ excess, while $\kappa\gtrsim 7-12$ will result in a $5\sigma$ excess in the combined
$t\bar{t}h(h\rightarrow\mu\mu)$ channels and the clear discovery of a beyond-the-Standard-Model Higgs boson.   

\section{ Conclusions}
 
In this letter, we have investigated the observability of a light, CP-even Higgs boson at the LHC in the channel $t\bar{t}h(h\rightarrow\mu\mu)$.  We have shown that for a light SM Higgs boson with a mass around 120~GeV, evidence for a 
Standard Model Higgs can be obtained in this channel at roughly the same significance level as 
in the $gg \rightarrow h\rightarrow \mu\mu$ and $qq'\rightarrow qq'h(h\rightarrow\mu\mu)$
channels, and with comparable Higgs-mass resolution.   The best statistical significance is obtained in the fully hadronic mode.  Moreover, we have shown that for $m_h$ between 115~GeV and 130~GeV, the combined results from the
gluon-fusion, weak-boson fusion, and $t\bar{t}h$ channels, with $h$ decaying to $\mu^+\mu^-$,
can provide evidence for a SM Higgs boson at about $3\sigma$ or better  with $300\ {\rm fb}^{-1}$ of integrated luminosity.  In beyond-the-Standard-Model scenarios in which the 
$t\bar{t}h$ production cross-section and/or the branching fraction $\mathrm{BR}(h\rightarrow\mu\mu)$ are enhanced relative to their SM values, this 
channel could become extremely important for the detection and analysis of a light Higgs 
boson, providing an accurate measurement of the Higgs mass, and a precise determination 
of the muon Yukawa coupling.  


\section{Acknowledgments}

We wish to thank  A. Belyaev,
T.~Han, B.~McElrath, and M.~Schmitt for correspondence and discussion.
We also wish to thank the Institute for Nuclear Theory at the University of Washington for its hospitality and the Department of Energy for partial support during the completion of this work.
This work was supported in part
by the Department of Energy under Grant~DE-FG02-04ER-41298.


\smallskip


\end{document}